\documentstyle[11pt,epsf]{article}

\newfont{\subsub}{cmr6}

\newcounter{szk}

\begin{document}
\title{Fractal Structure with a Typical Scale
}
\author{
\footnote{e-mail address: anazawa@tohtech.ac.jp} Masahiro Anazawa$^a$, 
\footnote{e-mail address: ishikawa@kanazawa-gu.ac.jp} Atushi Ishikawa$^b$, 
\footnote{e-mail address: tadao@nanao-c.ac.jp} Tadao Suzuki$^c$ and
\footnote{e-mail address: mtomo@curie.sci.u-ryukyu.ac.jp} 
Masashi Tomoyose$^d$\\ \\
$^a$ Tohoku Institute of Technology, Sendai 982-8577, Japan\\
$^b$ Kanazawa Gakuin University, Kanazawa 920-1392, Japan\\
$^c$ Nanao Junior College, Nanao 926-8570, Japan\\
$^d$ University of the Ryukyus, Nishihara 903-0213, Japan
}
\date{}
\maketitle

\begin{abstract}
In order to understand characteristics common to distributions which have both 
fractal and non-fractal scale regions in a unified framework, we introduce a 
concept of typical scale.
We employ a model of 2d gravity modified by the $R^2$ term as a tool to 
understand such distributions through the typical scale. 
This model is obtained by adding an interaction term with a typical scale to a 
scale invariant system.
A distribution derived in the model provides power law one in the large 
scale region, but Weibull-like one in the small scale region.
As examples of distributions which have both fractal and non-fractal regions, 
we take those of personal income and citation number of scientific papers.
We show that these distributions are fitted fairly well by the distribution 
curves derived analytically in the $R^2$ 2d gravity model.
As a result, we consider that the typical scale is a useful concept to 
understand various distributions observed in the real world in a unified way.
We also point out that the $R^2$ 2d gravity model provides us with an 
effective tool to read the typical scales of various distributions in a 
systematic way.
\end{abstract}


\vspace{1cm}
\section{Introduction}
\indent

A self-similar system is called fractal \cite{Mandelbrot} and it is one of the 
subjects which attract attention broadly not only in natural science but also 
in social science, in recent years.
In many cases, fractal structure appears in some restricted scale, does not do 
in all scale of the system concerned.
For example, in the case of distribution of personal income interested in 
econophysics \cite{MS}, the distribution of top several percent income 
earners follows fractal power law \cite{Pareto}, while that of the rest earners 
does not do \cite{Gibrat, Badger, ASNOTT}.  
In other words, in such a system, self-similarity is not maintained over all 
scale but is broken in small scale region. 

In usual, the fractal property and the deviation from it in each system are 
discussed by using individual models for the system.
For example, as for the personal income distribution, the fractal power law 
in high income region is studied using a stochastic evolution equation 
\cite{LS}.
In addition, whole profile of the distribution is investigated based on a 
concept of small world network \cite{SFA}, and in \cite{Borges}, the 
distribution is explained using q-Gaussian distribution emerged from 
nonextensive statistical mechanics \cite{Tsallis}.
In this paper, however, we will discuss characteristics common to systems 
which show both fractal and non-fractal properties independent of the detail 
of the individual models.

A scale invariant model does not have any scale, so we expect that 
distributions derived from the model follow power law.
This is thought to be a universal property that does not depend on the 
detailed structure of the model.
One of the simplest methods to realize both fractal and non-fractal scale 
regions analytically is to introduce a typical scale into a scale invariant 
model to break the original scale invariance.
If the interaction term with a typical scale is added to the scale invariant 
model, the model obtains a typical scale and we expect that the distributions 
derived form the model become non-fractal in the scale region where the 
typical scale is meaningful, while keeping fractal property in the large scale 
region.
In this paper, we will discuss fractal and deviation from it in the above 
framework.

There are various ways to construct a scale invariant model. As a tool to 
discuss fractal property and deviation from it concretely, we take a model of 
2-dimensional quantum gravity coupled with conformal matter fields.
Because of special nature of 2-dimension and the conformal property of the 
matter fields, the model has scale invariance and fractal property. 
Some distributions derived from the model follow power law\cite{BIPZ, 
KAD}.
One of the reasons to take this model is that the model is decided by the 
action only and it is suitable to treat the whole system analytically. 
In addition, the model has a simple geometrical meaning, so that it is easy to 
understand the fractal property and the deviation from it 
intuitively.\footnote{
It is believed that a typical 2d surface has self-similar
structure (Fig.~\ref{fig:2-dim Random Surface}).
}

The above model, the standard 2d gravity model coupled with conformal 
matter fields, is scale invariant.
To introduce a typical scale and break the original scale invariance, we add 
the $R^2$ interaction term to the action. Here, $R^2$ is the square of scalar 
curvature.  The obtained model \textbf{is} called $R^2$ 2d gravity theory.
Because of the typical scale introduced, we expect the deviation from fractal 
in the scale region where the typical scale is meaningful\cite{KN, ITY}, 
keeping the original fractal property in the large scale region. 

In this paper, we would like to point out that the typical scale is a useful new 
concept to understand various distributions which have both fractal and 
non-fractal regions, employing the $R^2$ 2d gravity model as a tool to 
understand the features of such distributions in a unified way. 
As examples of distributions which have fractal and non-fractal regions, we 
take those of personal income and citation number of scientific papers.
We show that these distributions are well understood 
by the typical scales and theoretical curves derived from
the framework of the $R^2$ 2d gravity model.
We also point out that the $R^2$ 2d gravity model also provides us with an 
effective tool to read the typical scales of various distributions in a 
systematic way.

\section{2-dimensional gravity with $R^2$ term}
\label{sec-R2Gin2D}
\indent

In this section, we review $R^2$ 2d gravity model.
First let us consider standard 2-dimensional quantum gravity coupled
with conformal matter fields. 
To make the argument concrete, as conformal matter fields we 
take scalar fields  $X^i(i=1,2,\cdots ,c)$.
The action of the matter part takes the form
\begin{eqnarray}
    S_M(X^i;g)=\frac{1}{8 \pi} \int {\rm d^2}x \sqrt{g} g^{\mu \nu} 
    \partial_{\mu} X^i \partial_{\nu} X^i~,
    \label{action X}
\end{eqnarray}
where $g_{\mu \nu}(\mu, \nu=0, 1)$ is the metric of 2d surface.
In 2-dimension, the standard Einstein action 
$
    \frac{1}{4 \pi} \int {\rm d^2}x \sqrt{g}\, R \;,
$
where $R$ is the scalar curvature, merely yields a constant which
characterizes the topology of the 2d surface, so that we can neglect the
Einstein term. 
The total action is given by 
$
    S_{total}(X^i;g) = S_M(X^i;g) \, ,
$
and it is invariant under the scale transformation of
the metric $g_{\mu \nu}$.

The partition function for fixed area $A$ of 2d surface 
is given by
\begin{eqnarray}
    Z(A)=\int \frac{{\cal D} g 
    {\cal D} X}{\rm vol(Diff)}
    {\rm e}^{-S_{total}(X^i;g)}~ \delta(\int d^2 x \sqrt{g}-A)~.
    \label{partition}
\end{eqnarray}
The action and the integration measure ${\cal D}g {\cal D}X$
are invariant under 2d diffeomorphisms, so that
the measure should be divided by the volume of the 
diffeomorphisms, which is denoted by $\rm vol(Diff)$.
The partition function is  evaluated to be
$
    Z(A) \propto A^{\gamma_{\infty }-3} 
$
\cite{BIPZ},
where $\gamma_{\infty}$ is a constant determined by the central
charge $c$ and the number of handles of the 2d surface $h$,
\begin{eqnarray}
    \gamma_{\infty }(c, h) = \frac{c-25-\sqrt{(25-c)(1-c)}}{12}(1-h)+2~.
\label{gamma_infty}
\end{eqnarray}
Note that this model has no scale parameter, so that the partition 
function $Z(A)$ follows power law. It is expected that a typical
2d surface has self-similar structure (Fig.~\ref{fig:2-dim Random
Surface})
\cite{BIPZ}.

Next let us turn to 2d gravity with $R^2$ term.
In order to introduce a typical scale into the standard 2d gravity,
we add the scale variant $R^2$ term \cite{KN}
$
    1/(32\pi m^2) \int {\rm d^2}x \sqrt{g} R^2
$
to the action (\ref{action X}).
The total action is given by
\begin{eqnarray}
    S_{total}(X^i;g)=\frac{1}{32 \pi m^2} \int {\rm d^2}x \sqrt{g}\, R^2
      + S_M(X^i;g) \;,
    \label{action R2}
\end{eqnarray}
where $m$ is a coupling constant of length dimension $-1$.
The first term in the action (\ref{action R2}) is not scale invariant, 
and it provides the typical scale $2\pi/m^2$ to the theory. 
As a result, the fractal structure of 2d surface collapses in the region
where the typical scale $2\pi/m^2$ becomes meaningful, while 
2d surface maintains fractal at area scale much larger than 
$2\pi/m^2$.
In fact, the asymptotic forms of the partition function
are evaluated as \cite{KN}
\begin{eqnarray}
Z(A)
&\sim& C_0~A^{\gamma_0-3}~\exp\biggl[
                    -\frac{2 \pi}{m^2 A}(1-h)^2 \biggr]
                            ~~~~~~~~~~{\rm for}~~A \ll \frac{2\pi}{m^2}~,
\label{0}
\\
&\sim& C_{\infty}~A^{\gamma_{\infty }-3}
                            ~~~~~~~~~~~~~~~~~~~~~~~~~~~~~~~~~~~~~
                            {\rm for}~~ A \gg \frac{2\pi}{m^2}~,
\label{infty}
\end{eqnarray}
where $C_0$ and $C_\infty$ are the proportional constants, and
\begin{eqnarray}
    \gamma_0(c, h)=\frac{(c-12)}{6}(1-h) + 2~ \;.
\label{gamma_0}
\end{eqnarray}
Here we can observe that fractal power law is broken 
in the region $A \ll 2\pi/m^2$.

In order to investigate the breaking of fractal structure concretely,
it is appropriate to treat 2d surface discretely. 
In study of 2d gravity, 
one of the useful methods of discretizeing
2d surface is known as 
Dynamical Triangulation (DT) \cite{KAD}. 
In usual, 2d surface is discretized using small equilateral triangles,
where each triangle has the same size. From various evidences,
DT is believed to be equivalent to the continuum theory of 2d gravity
in the continuum limit \cite{Takusan}.

In DT, the evaluation of the partition function is performed 
by replacing the path integral over the metric with the sum over 
possible triangulations of 2d surface.
Here, we represent the number of triangles sharing the vertex $i$ as $q_i$,
which is called a coordination number.
In DT, the $R^2$ term in the action
(\ref{action R2}) is expressed by
\begin{eqnarray}
    \int d^2 x \sqrt{g} R^2  
    \cong \frac{4 \pi^2}{3 a^2}  \sum_i 
      \frac{(6-q_i)^2}{q_i} ~, 
\label{discretized R2}
\end{eqnarray}
from the correspondence $\int d^2 x \sqrt{g} \cong a^2 \sum_i q_i /3$ 
and $R_i = 2\pi (6-q_i)/(a^2 q_i)$.
Here $a^2$ is the area of a triangle 
and $R_i$ is the discretized local scalar curvature at the $i$-th
vertex. From Eq.~(\ref{discretized R2}), we can recognize that the $R^2$ term
has the effect 
to make 2d surface flat ($q_i=6$).
This effect is parametrized by the coefficient of the $R^2$ term.

\section{MINBU distribution}
\label{sec-minbu}
\indent

\begin{figure}[htb]
 \begin{minipage}{0.43\textwidth}
  \begin{center}
   \epsfysize=50mm
   \leavevmode
   \epsfbox{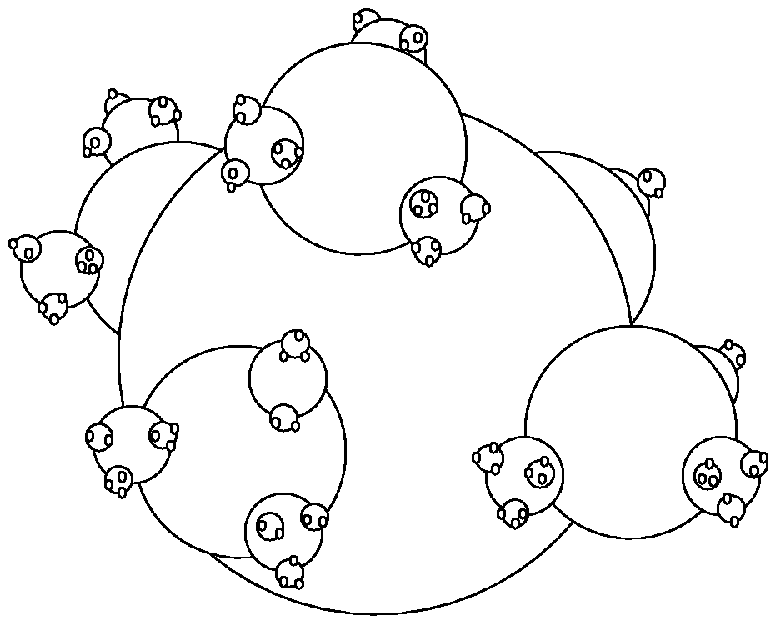}
  \end{center}
  \vspace{-5mm}
  \caption{A typical 2-dim surface.}
  \label{fig:2-dim Random Surface}
 \end{minipage}
 \begin{minipage}{0.55\textwidth}
  \begin{center} 
   \epsfysize=50mm
   \leavevmode
   \epsfbox{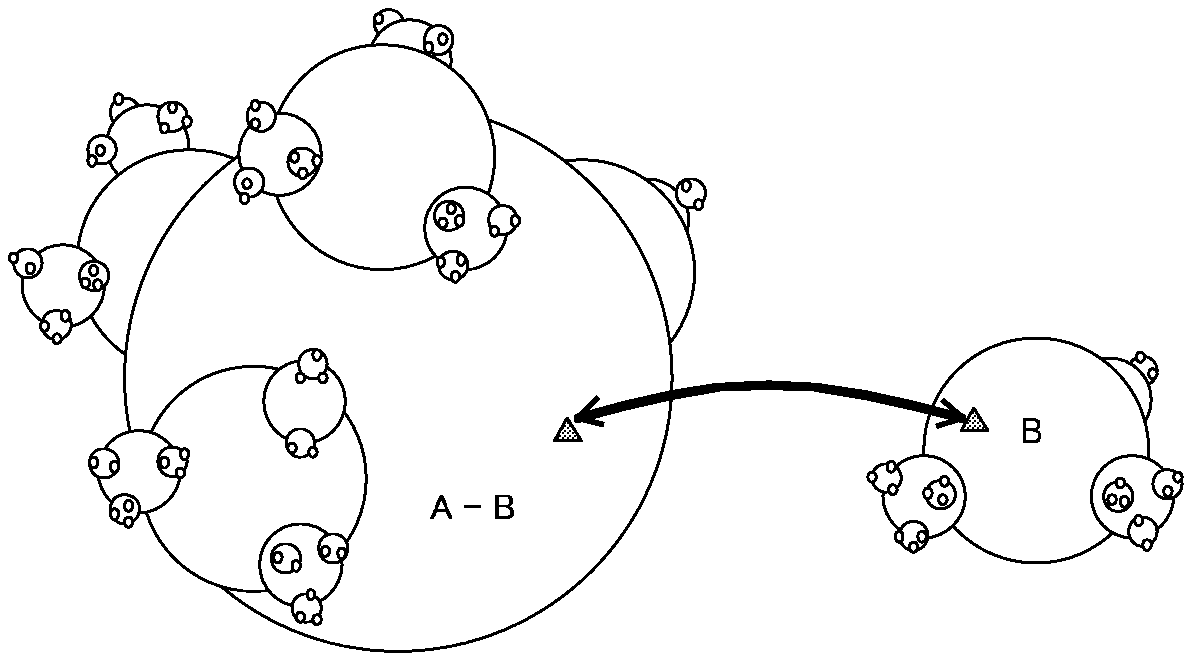}
   \caption{Divided MINBs.}
   \label{fig:Divided MINBs}
  \end{center}
 \end{minipage}
\end{figure}

In DT, fractal structure (and non-fractal structure) of 2d surface
can be discussed by considering  so-called minimum-neck baby universe 
(MINBU) \cite{JM}.
A MINBU is defined as a simply connected area region of 2d surface
whose neck is composed of three links
(three sides of triangles),
where the neck is closed and non-self intersecting.
In general, a lot of MINBUs of various sizes are formed
on a 2d surface.
A typical dynamically triangulated surface is shown in 
Fig.~\ref{fig:2-dim Random Surface}.  
Distribution of the area of MINBU is one of the
important observable quantities in DT. 

Now let us evaluate the distribution of MINBU.
Consider a closed 2d surface of area $A$.
There are many MINBUs on the surface, and
each one is connected by a minimum neck one another.
Paying attention to one of the minimum necks, the whole
surface can be divided into two MINBUs (Fig.~\ref{fig:Divided MINBs}),
where one has area $A-B$ and the other has area $B$.
Representing the partition functions of the two MINBUs as $Z(A-B,3)$
and $Z(B,3)$ respectively, the statistical average number of
finding a MINBU of area $B$ on a closed surface of area $A$,
$n_A(B)$,  can be expressed as 
\begin{eqnarray}
 n_A(B)\sim \frac{Z(B, 3) Z(A-B, 3)}{Z(A)}~.
\label{n_A(B)}
\end{eqnarray}
Here we set $a^2=1$ for simplicity, and $Z(A)$ denotes the partition
function of a closed surface of area $A$.

On the other hand, a MINBU of area $C$ can be constructed by removing
one triangle 
from a closed 2d surface of area $C+1$. We have $C+1$ ways to choose the
triangle to remove, so we have the relation 
\begin{eqnarray}
 Z(C, 3) \sim (C+1)Z(C+1)~.
\end{eqnarray}
Using this relation, the partition functions of the two MINBUs in
Eq.~(\ref{n_A(B)}) are given by
\begin{eqnarray}
 Z(B, 3)&\sim& (B+1)Z(B+1)~,
 \label{B}
\\
 Z(A-B, 3)&\sim& (A-B+1)Z(A-B+1)~.
 \label{A-B}
\end{eqnarray}
Substituting these relations into Eq.~(\ref{n_A(B)}), we can express $n_A(B)$
in terms of the ordinary partition functions of closed surfaces.
Using the asymptotic forms of the partition function (\ref{0}) and 
(\ref{infty}),
we obtain, in the end, the asymptotic expression of $n_A(B)$
\begin{eqnarray}
 n_A(B)
&\sim& C_0~A~ 
      B^{\gamma _0 -2}
      \exp\biggl[ -\frac{2\pi}{m^2B}(1-h)^2 \biggr]
\qquad\qquad
      {\rm for }~~ 1 \ll B \ll \frac{2\pi}{m^2} \ll A~,
 \label{eq-n_A-low}\\
&\sim& C_{\infty}~A~  \biggl[(1-\frac{B}{A})B \biggl]^{\gamma _\infty -2}
\qquad\qquad\qquad~~~~~~
      {\rm for }~~\frac{2\pi}{m^2} \ll B < A/2~.
 \label{eq-n_A-hight}
\end{eqnarray}
Here, the restriction $B < A/2$ comes from the strict definition
of MINBU, where the area of a MINBU is less than half of the total area.

As for the case $2\pi/m^2 \ll B < A/2$,
the asymptotic form
(\ref{eq-n_A-hight})
follows power law, therefore,
the surfaces are expected to be fractal
\footnote{A similar phenomenon can be seen in 2d gravity 
without the $R^2$ term \cite{BIPZ},
where no typical length scale exists.}.
In this range, even if the model contains the $R^2$
term, at an area scale much larger than $2\pi/{m^2}$, the surfaces are 
fractal. 
On the other hand, as for the case $1 \ll B \ll 2\pi/m^2$,
the asymptotic form
(\ref{eq-n_A-low}) is highly suppressed by the
exponential factor $\exp\bigl[ -\frac{2\pi}{m^2B}(1-h)^2 \bigr]$, 
hence, the fractal structure of 2d surface is broken.
In this range,  
at an area scale much smaller than $2\pi/{m^2}$, the surfaces
are affected by the typical length scale, and are not fractal.
In the case of $\gamma_0 = 0$,
the distribution (\ref{eq-n_A-low}) is known as Weibull distribution.
In this paper, we call the distribution (\ref{eq-n_A-low})
as Weibull-like distribution.

\section{Numerical analysis of DT}
\label{sec-parameters}
\indent

\begin{figure}[htb]
 \centerline{\epsfxsize=0.9\textwidth\epsfbox{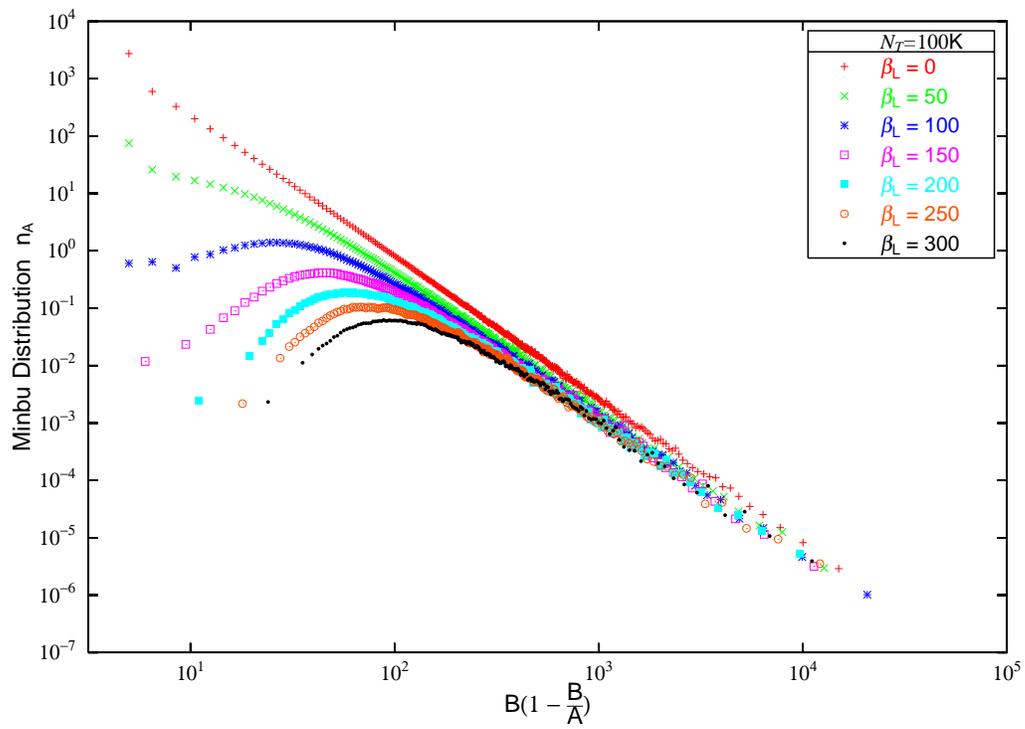}}
 \caption{The simulation results of DT.}
 \label{fig:The Simulation Results of DT}
\end{figure}

\begin{figure}[htb]
 \centerline{\epsfxsize=0.78\textwidth\epsfbox{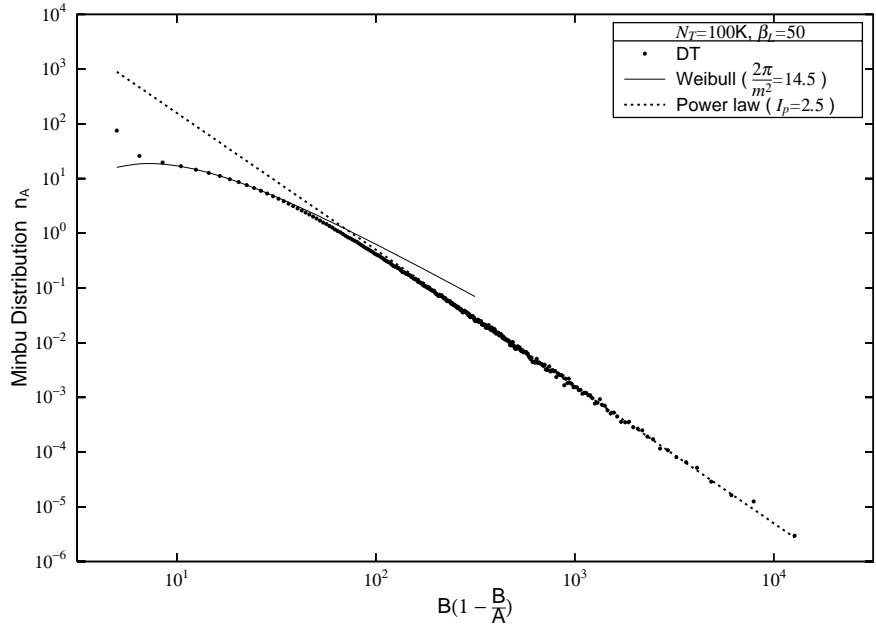}}
 \caption{The fitting of $\beta_L=50$ data.}
 \label{fig:beta=50}
\end{figure}

\begin{figure}[htb]
 \centerline{\epsfxsize=0.78\textwidth\epsfbox{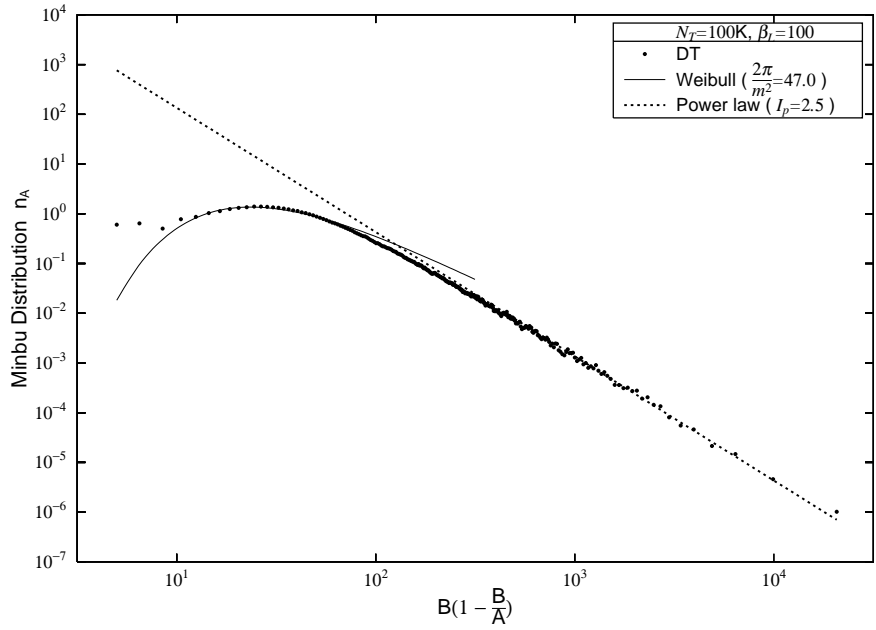}}
 \caption{The fitting of $\beta_L=100$ data.}
 \label{fig:beta=100}
\end{figure}

The analytic results (\ref{eq-n_A-low}) and (\ref{eq-n_A-hight}) can be
confirmed in the simulation of DT
for the simple case that 2d surface is sphere ($h=0$)
and there is no matter field on it ($c=0$) \cite{ITY}. 
The simulation results are expressed
in Fig.~\ref{fig:The Simulation Results of DT}.
Here, we plot MINBU distributions, $n_A(B)$ versus $(1-B/A)B$ 
with a log-log
scale for $\beta_L = 0$, $50$, $150$, $200$, $250$, $300$, which are 
coefficients of 
the discretized $R^2$ term (\ref{discretized R2}) \footnote{
In Eq.~(\ref{eq-n_A-low}), we can replace $B$ with $(1-B/A)B$
because of the range $B \ll 2\pi/{m^2} \ll A$.
}.
In this simulation, the total number of triangles is 100,000.
These MINBU distributions can be well explained 
by the asymptotic formulae (\ref{eq-n_A-low}) and (\ref{eq-n_A-hight})
with $\gamma_0=0$ and $\gamma_\infty=-1/2$,
which are obtained from $h=c=0$.
We can read the typical scale $2\pi/m^2$ for each case.
For example, the data fittings for the cases of 
$\beta_L=50$ and $100$ are represented  
in Figs.~\ref{fig:beta=50} and \ref{fig:beta=100},
and we obtain $14.5$  and $47.0$
as the value of $2\pi/m^2$ respectively.
In each of these figures, several data points for small MINBUs are
apart from the line of Weibull distribution (\ref{eq-n_A-low}).
We consider that it is the finite lattice effect.
In small $B$ region, each of the corresponding MINBUs consists of a
small number of triangles, so that it is not appropriate to treat
the area of MINBU $B$ as a continuous variable.

\section{Distributions of personal income and citation number of scientific papers}
\label{sec-fitting}
\indent

We apply the distribution of MINBU in 2d $R^2$ gravity
to other distributions observed in the real world,
and examine whether it can explain these distributions.
Here, we investigate distributions of personal income and 
citation number of scientific papers \cite{Render}.
These two kinds of distributions have fractal power law and non-fractal
regions, so it is possible that the theoretical curves (\ref{eq-n_A-low}) 
and (\ref{eq-n_A-hight}) can explain  them.

\begin{figure}[htb]
 \centerline{\epsfxsize=0.90\textwidth\epsfbox{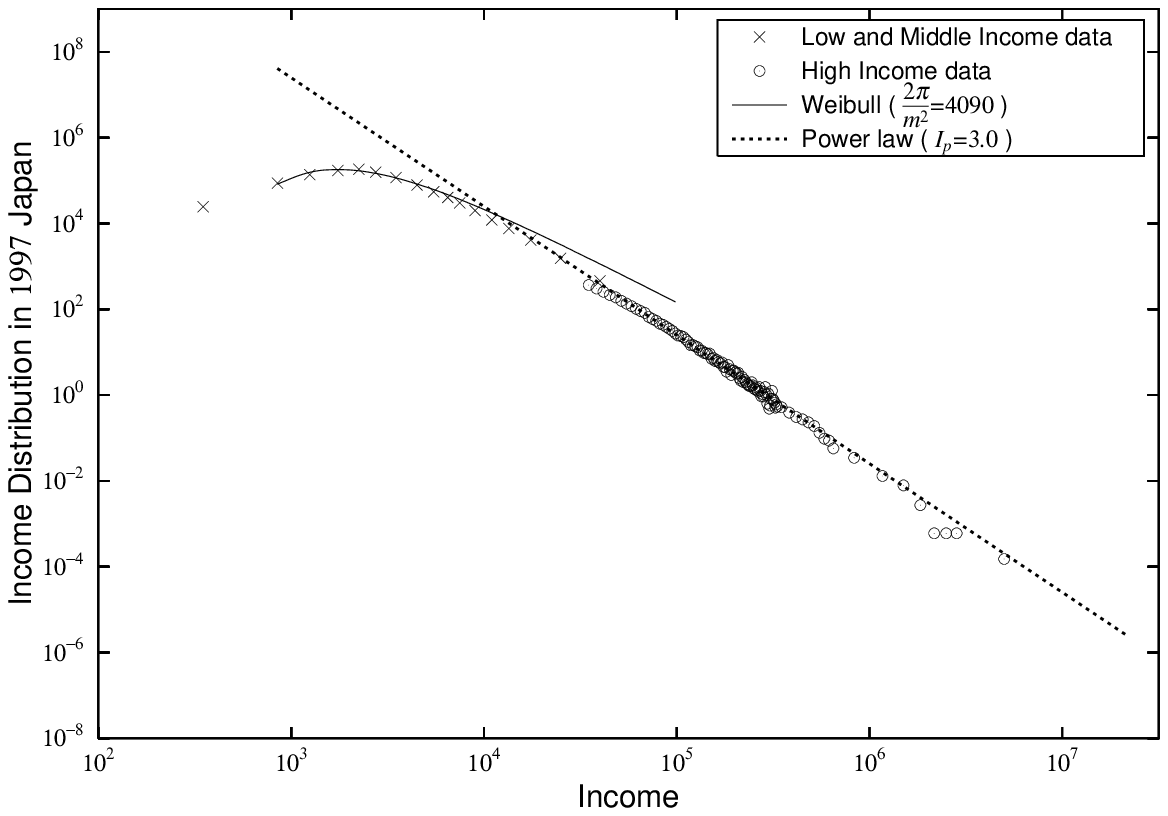}}
 \caption{The fitting of the personal income distribution in 1997 Japan.}
 \label{fig:1997-japan}
\end{figure}

\begin{figure}[htb]
 \centerline{\epsfxsize=0.90\textwidth\epsfbox{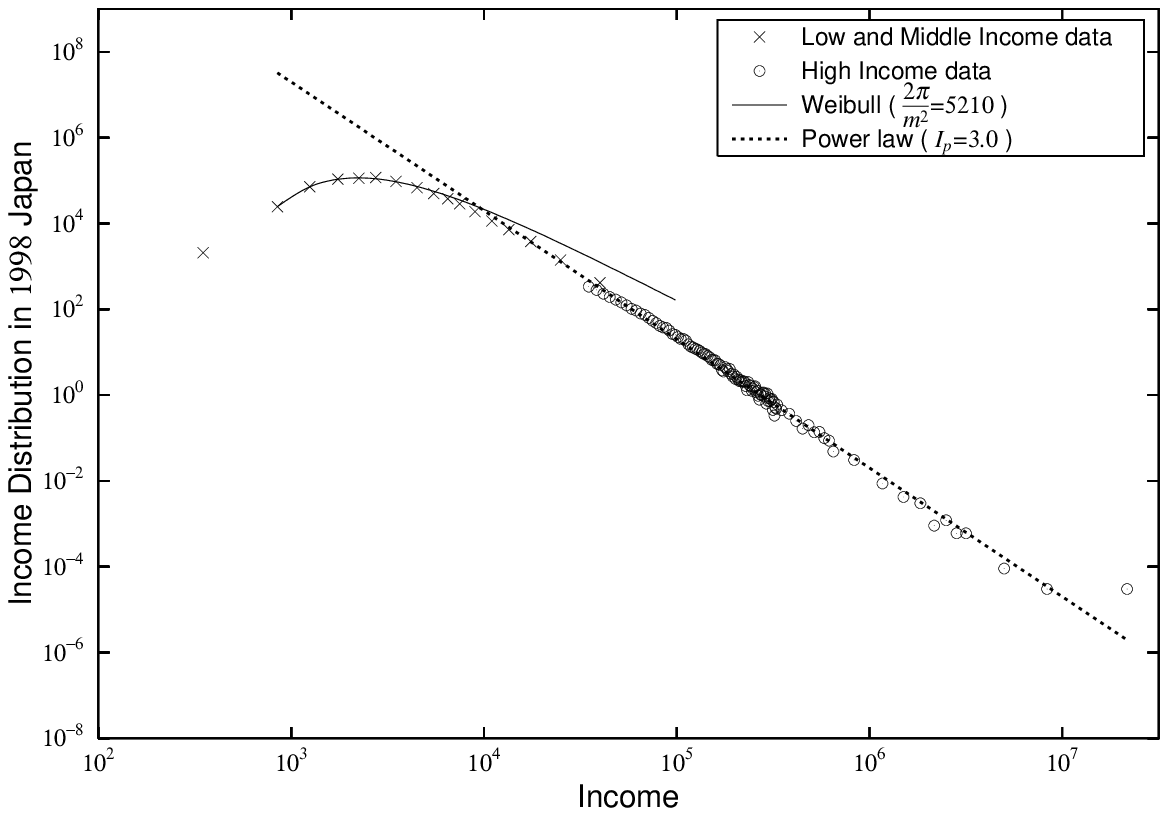}}
 \caption{The fitting of the personal income distribution in 1998 Japan.}
 \label{fig:1998-japan}
\end{figure}

First, let us consider the personal income distributions of Japan 
in the years 1997 and 1998 \cite{ASNOTT}.
The distributions and data fittings are shown in
Figs.~\ref{fig:1997-japan} and \ref{fig:1998-japan}.
Here, we do not accumulate the data in this analysis.
The horizontal axis indicates the income $x$ in units of thousand yen 
and the vertical axis indicates
the number density of persons $N(x)$ 
per a period of $100$  thousand yen.

In both distributions, from the data fittings for the power law regions,
we obtain $\gamma_\infty-2 \equiv  -I_p = -3.0$. From Eq.~(\ref{gamma_infty}),
we see that this value is realized by
choosing $c=-2$, $h=0$.
Substituting these values into Eq.~(\ref{gamma_0}), we obtain $\gamma_0 -
2 = -7/3$ 
for the Weibll-like distribution (\ref{eq-n_A-low}).
The analytical functions employed to fit the personal income distributions
are given by 
\begin{eqnarray}
   N(x)&\sim& C_{\rm w}~
   x^{-7/3}~
   \exp\biggl[ -\frac{2\pi}{m^2} \frac{1}{x}
   \biggr]
   ~~~~~~~~~~~~~~~~~   {\rm for}~~~~ 1 \ll x \ll \frac{2\pi}{m^2}~,
   \label{eq-n_A(B)W}
\\
   &\sim& C_{\rm p}~
   x^{-3}
   ~~~~~~~~~~~~~~~~~~~~~~~~~~~~~~~~~~~~~~
   {\rm for}~~~~ \frac{2\pi}{m^2} \ll x~.
   \label{eq-n_A(B)PL}
\end{eqnarray}
In Figs.~\ref{fig:1997-japan} and \ref{fig:1998-japan}, 
we fit the data in the non-fractal regions 
by the Weibull-like distribution~(\ref{eq-n_A(B)W}),
and find that the typical scales $2\pi/m^2$ in 1997-
and 1998-Japan are $4090$ and $5210$ thousand yen
respectively.
These values are almost the same as the 
averages of income, 
and
we consider that these values are quite natural.
We note that 
the scale transformation of $x$
and the adjustment of the normalization of $N(x)$
can always make the normalization constants 
$C_{\rm w}$ and $C_{\rm p}$ agree with
the corresponding constants 
$C_0$ and $C_\infty$ in Eqs.~(\ref{eq-n_A-low}) and
(\ref{eq-n_A-hight}) respectively.
In the end, we consider that these two distributions of personal income
are well explained by 
a concept of typical scale in distributions.

\begin{figure}[htb]
 \centerline{\epsfxsize=0.78\textwidth\epsfbox{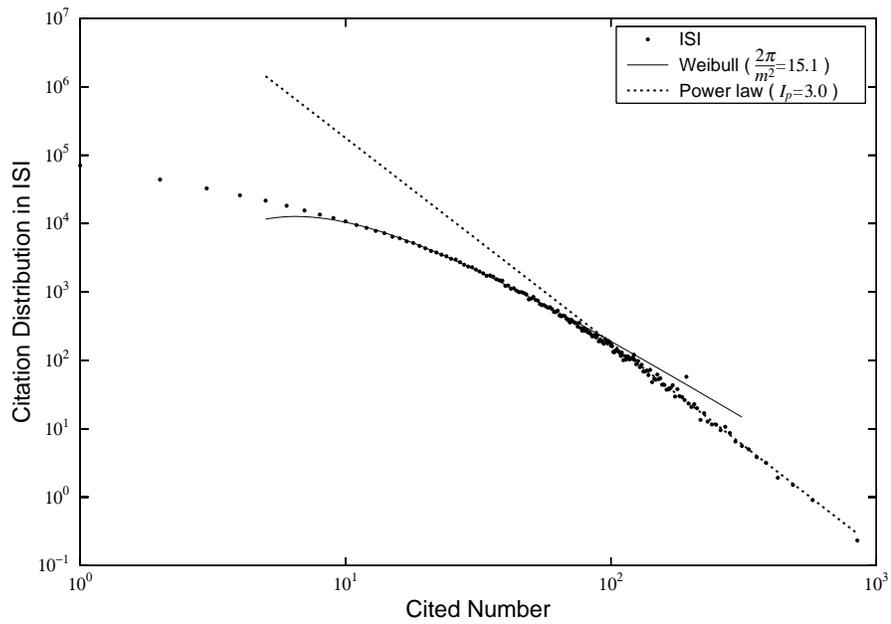}}
 \caption{The fitting of the citation number distribution in ISI.}
 \label{fig:ISI}
\end{figure}

\begin{figure}[htb]
 \centerline{\epsfxsize=0.78\textwidth\epsfbox{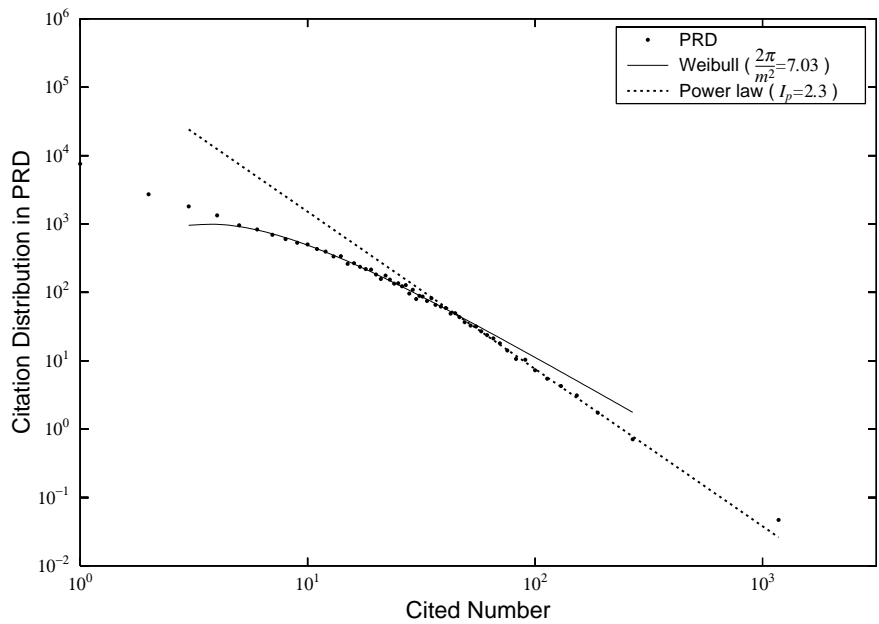}}
 \caption{The fitting of the citation number distribution in PRD.}
 \label{fig:PRD}
\end{figure}

Secondly, we consider two distributions of citation number  
of scientific papers analyzed in Ref.~\cite{Render}.
One is the citation number distribution of the papers
published in 1981 in journals which are 
cataloged by the Institute for Scientific
Information (ISI).
The second is that of the papers which
were published in vols.~11-50 of Physical
Review D (PRD), 1975-1994.
As for the data of ISI,
the distribution can be well explained 
by setting $c=-2$, $h=0$ (Fig.~\ref{fig:ISI}). 
On the other hand,
as for the data of PRD,
the distribution can also be  well explained
by setting $c=1/2$, $h=0$ (Fig.~\ref{fig:PRD}).
In the latter case, to fit the data, we employ the analytic functions
\begin{eqnarray}
   N(x)&\sim&  C_{\rm w}~
   x^{-23/12}~
   \exp\biggl[ -\frac{2\pi}{m^2} \frac{1}{x}
   \biggr]
   ~~~~~~~~~~~   {\rm for}~~~~ 1 \ll x \ll \frac{2\pi}{m^2}~,
   \label{citation}\\
   &\sim& C_{\rm p}~
   x^{-7/3}
   ~~~~~~~~~~~~~~~~~~~~~~~~~~~~~~~~
   {\rm for}~~~~ \frac{2\pi}{m^2} \ll x~.
\end{eqnarray}
We find that the typical scales $2\pi/m^2$ of the citation number
in ISI and PRD are $15.1$ and $7.03$ respectively.
In the small $x$ regions in Figs.~\ref{fig:ISI} and \ref{fig:PRD},
several data points are apart from the curves of the Weibll-like
distributions  (\ref{eq-n_A(B)W}) and (\ref{citation}).
In the derivation of the Weibull-like function, the area of
MINBU is treated as a continuous variable.
However, it is not appropriate to treat $x$ continuously
in small $x$ region.
As a result, we consider that $x$ does not always follow 
the Weibull-like distribution in small $x$ region.

\section{Summary and discussion}
\label{sec-summary}
\indent

In this paper, we proposed the concept of a typical scale in order to 
understand distributions which have both fractal and non-fractal scale 
regions in a unified framework.
The point was to introduce a typical scale into a scale invariant model to 
break the original scale invariance and to produce non-fractal feature in the 
small scale region.
We employed the $R^2$ 2d gravity model as a tool to understand such 
distributions through the typical scale.
MINBU distribution in this model followed the power law in the large scale 
region and provided Weibull-like one in the small scale region.

As examples of distributions where fractal and non-fractal regions coexist, 
we took those of personal income and citation number of scientific papers.
We showed that these distributions were fitted fairly well by the theoretical 
curves of MINBU distribution, adjusting the values of $c$, $h$ and the typical 
scale. From these fittings, as for the personal income,
we consider that there is no scale with respect to money for
the top several percent high income earners, on the other hand,
the rest earners are highly influenced by the typical scale of income.
We can understand whole profile of the distribution
merley by introducing the typical scale.
As a result, we consider that the typical scale is a useful concept to 
understand various distributions where both fractal and non-fractal scale 
regions exist. 
We also consider that the $R^2$ 2d gravity model provides us with an 
effective tool to read the typical scales of various distributions in a 
systematic way.

In the distributions studied in this paper, the values of the typical scale are 
comparable with the average values of the distributions. We consider that 
the typical scale is a significant characteristic parameter 
similar to average in such distributions. 
The typical scale, however, can be read mainly 
from the data in the small scale region. It can thus be a more efficient 
characteristic parameter than average in some situations.

We use the $R^2$ 2d gravity model as a tool to discuss the significance of the 
typical scale concretely. 
Besides the coincidence of the distributions, is there any direct physical 
connection between $R^2$ 2d gravity and personal income or citation number?
We can't answer this question at present.
However, 2d gravity can be also formulated by a stochastic evolution 
equation \cite{JR}.
As we mentioned in Sect.~1, personal income distribution can be also 
described by a stochastic evolution equation \cite{LS}.
It may be possible to find the physical relation by investigating these
formulations in detail.


\section*{Acknowledgements}
\indent

The authors would like to express our gratitude to Dr.~W. Souma,
Dr.~Y. Fujiwara, Professor.~H. Aoyama and 
Professor~H. Terao for valuable advices and discussions.  
We are also grateful to Professor~H. Kawai for useful comments
especially on his work \cite{KN}. 
Thanks are also due to members of YITP,
where one of the authors (A. I.) stayed several times during 
the completion of this work.



\newpage


\end{document}